\documentclass[sigconf]{acmart}

\usepackage{booktabs} 
\usepackage{listings}
\usepackage{tikz}
\usepackage{pgfplots}
\pgfplotsset{compat=newest}
\usepackage{pgfplotstable} 
\usepgfplotslibrary{patchplots}
\pgfplotscreateplotcyclelist{cbw}{%
    blue,           every mark/.append style={                     fill=blue!80!black   }, mark=*\\%
    green!50!black, every mark/.append style={                     fill=green!30!black  }, mark=square*\\%
    brown!60!black,densely dotted, every mark/.append style={                     fill=brown!80!black  }, mark=triangle*\\%
    black,                                                                                 mark=star\\%
    red,            densely dashed,every mark/.append style={solid,fill=red!80!black    }, mark=*\\%
                    densely dashed,every mark/.append style={solid,fill=brown!80!black  }, mark=square*\\%
    black,          densely dashed,every mark/.append style={solid,fill=gray            }, mark=otimes*\\%
    blue,           densely dashed,every mark/.append style={solid                      }, mark=star\\%
    red,            densely dashed,every mark/.append style={solid,fill=red!80!black    }, mark=diamond*\\%
}

\setcopyright{rightsretained}

\copyrightyear{2019} 
\acmYear{2019} 
\setcopyright{licensedusgovmixed}
\acmConference[ICPP 2019]{48th International Conference on Parallel Processing}{August 5--8, 2019}{Kyoto, Japan}
\acmBooktitle{48th International Conference on Parallel Processing (ICPP 2019), August 5--8, 2019, Kyoto, Japan}
\acmPrice{15.00}
\acmDOI{10.1145/3337821.3337906}
\acmISBN{978-1-4503-6295-5/19/08}

\begin{document}
\title{Automatic Differentiation for Adjoint Stencil Loops}

\author{Jan H\"uckelheim}
\orcid{0000-0003-3479-6361}
\affiliation{%
  \institution{Imperial College London}
  \streetaddress{South Kensington}
  \city{London}
  \country{United Kingdom}
  \postcode{SW7 2AZ}
}
\email{j.hueckelheim@imperial.ac.uk}

\author{Navjot Kukreja}
\affiliation{%
  \institution{Imperial College London}
  \streetaddress{South Kensington}
  \city{London}
  \country{United Kingdom}
  \postcode{SW7 2AZ}
}
\email{n.kukreja@imperial.ac.uk}

\author{Sri Hari Krishna Narayanan}
\affiliation{%
  \institution{Argonne National Laboratory}
  \streetaddress{9700 South Cass Road}
  \city{Lemont}
  \state{Illinois}
  \postcode{60439}
}
\email{snarayan@mcs.anl.gov}

\author{Fabio Luporini}
\affiliation{%
  \institution{Imperial College London}
  \streetaddress{South Kensington}
  \city{London}
  \country{United Kingdom}
  \postcode{SW7 2AZ}
}
\email{f.luporini12@imperial.ac.uk}

\author{Gerard Gorman}
\affiliation{%
  \institution{Imperial College London}
  \streetaddress{South Kensington}
  \city{London}
  \country{United Kingdom}
  \postcode{SW7 2AZ}
}
\email{g.gorman@imperial.ac.uk}

\author{Paul Hovland}
\affiliation{%
  \institution{Argonne National Laboratory}
  \streetaddress{9700 South Cass Road}
  \city{Lemont}
  \state{Illinois}
  \postcode{60439}
}
\email{hovland@mcs.anl.gov}

\renewcommand{\shortauthors}{J. H\"uckelheim et al.}

\begin{abstract}
Stencil loops are a common motif in computations including convolutional neural networks, structured-mesh solvers for partial differential equations, and image processing. Stencil loops are easy to parallelise, and their fast execution is aided by compilers, libraries, and domain-specific languages.
Reverse-mode automatic differentiation, also known as algorithmic differentiation, autodiff, adjoint differentiation, or back-propagation, is sometimes used to obtain gradients of programs that contain stencil loops. Unfortunately, conventional automatic differentiation results in a memory access pattern that is not stencil-like and not easily parallelisable.

In this paper we present a novel combination of automatic differentiation and loop transformations that preserves the structure and memory access pattern of stencil loops, while computing fully consistent derivatives. The generated loops can be parallelised and optimised for performance in the same way and using the same tools as the original computation.
We have implemented this new technique in the Python tool \texttt{PerforAD}, which we release with this paper along with test cases derived from seismic imaging and computational fluid dynamics applications.
\end{abstract}

%
%



\maketitle

\section{Introduction}
\begin{figure}
    \centering
    \def\svgwidth{1.0\linewidth}
    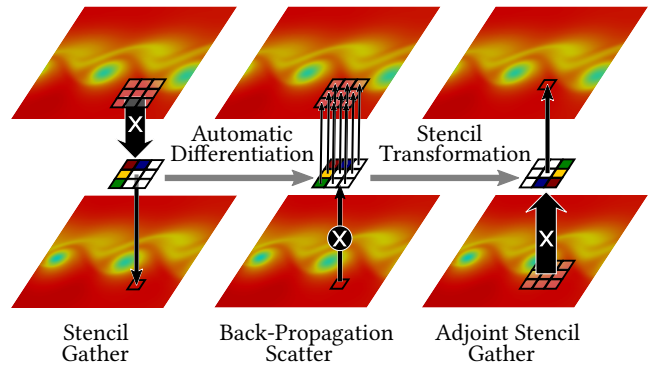
    \caption{Stencils with gather access occur in convolutions and many other applications and are easy to parallelise. Automatic differentiation or back-propagation transforms this operation into a scatter update, which is hard to parallelise without race conditions. With adjoint stencils, back-propagation can be performed by using only easily parallelisable stencils with gather access.}
    \label{fig:title}
\end{figure}

Derivatives are an important ingredient for optimisation, inverse modelling, error analysis, and related methods. Reverse-mode automatic differentiation (AD) is used to compute these derivatives in applications including climate modelling~\cite{heimbach2005efficient}, fluid dynamics~\cite{giles2005using}, machine learning~\cite{baydin2018automatic}, and image processing~\cite{li2018differentiable}. Starting from an implementation of a differentiable function, referred to as \emph{primal}, AD generates a new program that computes the derivative of that function. This is achieved by using operator overloading~\cite{griewank1996algorithm,hogan2014fast}, source-to-source~\cite{griewank1989automatic,utke2008openad,narayanan2010adic2,tapenade}, or just-in-time compilation~\cite{revels2016forward,innes2018don}.

In this paper, we focus on the differentiation of \emph{stencil loops}, which appear, for example, in convolutional neural networks and structured-mesh PDE solvers. Stencil loops are characterised by a memory access pattern where each index in an output array is updated based on data gathered from a neighbourhood around that index in one or more input arrays. Stencil computations are straightforward to parallelise. For example, when OpenMP is used, each thread is guaranteed to have a different value for the counter in the parallel loop, and hence the write operations will occur at unique memory locations. The read operations, on the other hand, occur at neighbourhoods around that index, and more than one thread can read data from the same memory location. Techniques to execute stencils efficiently on multicore CPUs or GPUs have been studied extensively; see for example~\cite{kamil2010auto,ragan2013halide}.

We aim to apply reverse-mode automatic differentiation by source-to-source transformation to such stencil loops. We will refer to this approach as AD for the remainder of this work. AD, also known as back-propagation or adjoint differentiation, allows the computation of gradients with respect to an arbitrary number of function inputs, at a cost that is independent of the number of inputs. Thus, one can generate high-resolution seismic images, perform industrial-scale shape optimisation, or train large neural networks. The derivatives generated by AD are also sometimes called \emph{adjoint} programs.

Reverse-mode AD traces derivatives backwards through a program, from the outputs to the inputs. In other words, the adjoint program is reading data from variables corresponding to the output of the primal program and is writing data to variables corresponding to the inputs of the primal.
This \emph{data flow reversal} also means that a concurrent read access in the primal program (i.e., a gather operation) can result in a concurrent write access in the derivative program (i.e., a scatter operation).
Consequently, a parallelisation or vectorisation strategy that is valid for the primal program may not be valid for its derivative, and reverse-differentiating parallel programs without introducing data races (while maintaining parallel efficiency) is a hard problem. Previous approaches work only for a small class of well-structured programs where array indices or pointer targets of input variables can be statically shown to be non-overlapping~\cite{forster2014algorithmic,ssmp}.

To avoid data races and therefore undefined program behaviour, one could safeguard every potentially conflicting write access with \texttt{atomic} pragmas or \texttt{critical} sections. Doing so, however, causes the write updates to be sequentialised, reducing parallel efficiency. Additionally, such constructs incur overheads due to the necessary thread synchronisation. The cost of executing the atomic updates may slow the program execution significantly, as we demonstrate in the test cases in this paper.

Another option is to use sum-reductions as implemented, for example, in the OpenMP runtime. In nontrivial scenarios, these require a private copy of the output array for each thread, which significantly increases the memory footprint of the program. Other approaches include colouring schemes, which incur an overhead in computing the colouring, as well as introducing synchronisation barriers for each colour, thus reducing parallel performance.

Our \emph{adjoint stencils} technique solves this problem by implementing back-propagation using only gather operations obtained via loop transformations of the scatter operator. It requires no additional memory and performs the bulk of the computation with no additional synchronisation barriers. All transformations are applied at compile time. The idea is illustrated in Figure~\ref{fig:title}.

The paper is organised as follows. We summarise related work and background in Section~\ref{sec:related}.
Following this, we describe the steps to generate adjoint stencils in Section~\ref{sec:method} and present test cases in Section~\ref{sec:testcases}. We present experimental performance results in Section~\ref{sec:results}, and in Section~\ref{sec:conclusion}, we summarise our conclusions and briefly describe future work.

\section{Related Work}
\label{sec:related}
AD differs from \emph{numerical differentiation} or \emph{finite differences} in that it computes exact derivatives of the implemented computation, with no truncation errors and without the need to choose any finite step size. Automatic differentiation also differs from symbolic differentiation in that it can handle large computations including loops, branches, and function calls efficiently.

AD for parallel programs has been addressed for distributed-memory MPI programs~\cite{hovland1997automatic}, exploiting the fact that message passing between parallel instances is explicit in the program source code and often limited to a relatively small, carefully defined interface. 
In this work, we focus on stencils, and not any specific parallelisation technique. Stencil compilers (e.g., YASK) can parallelise in MPI or shared memory but need the stencil structure.

Our approach is similar to that of autodiff for Halide~\cite{li2018differentiable}. For simple stencils, we expect both approaches to yield identical results. However, whereas the approach in~\cite{li2018differentiable} is intimately tied to Halide and relies on zero padding and sum reductions, our approach can be applied to any stencil compiler, such as Devito~\cite{devito-compiler},  HOSTS~\cite{stock2014framework}, and {\sc Tiramisu}~\cite{baghdadi2019tiramisu}, and requires neither padding nor reductions.

While our work uses some concepts from polyhedral compilers, it does so in a targeted way that is tailored to generating stencil loops that compute adjoints. Our framework supports nonrectangular iteration spaces, avoiding some of the overapproximation required in an interval-based representation like Halide.

A core aspect of our work is a gather-scatter conversion, similar to that described in~\cite{stock2014framework}. That paper, however, deals only with performance optimisation of stencil operations, and AD is not considered. To work with AD, the transformation must recognise that some inputs of the original function, as well as inputs to the adjoint computation, need to be gathered from the appropriate indices.

In a separate paper~\cite{tfmad}, we describe \emph{Transposed Forward-Mode AD} (TF-MAD) to transform the adjoint of a stencil back into a stencil loop. We applied this to a structured PDE solver and observed equally good scalability for the primal and adjoint. While that paper has an aim similar to the work presented here, it did not provide an implementation to automate the necessary transformations, and it was restricted to stencils with a symmetric data flow; that is, if the output array at index $i$ depends on the input array at index $j$, then the output at $j$ must also depend on the input at $i$. This is not the case for most stencil loops in practice, where the input space is slightly larger than the output space because of boundary effects.

Other authors have presented automatic differentiation for various domain-specific languages~\cite{farrell2013automated,paszke2017automatic}. These have in common that the differentiation needs to operate only on a high level of abstraction and assemble hand-optimised building blocks that perform efficient adjoint operations, for example, on neural network layers, partial differential equations, or linear algebra operators. In contrast, our work is not tied to any particular application domain and can differentiate any computation that has a stencil-like structure.

\section{Method and Implementation}
\label{sec:method}
In this section we describe the procedure to generate adjoint stencils and its implementation in the stencil differentiation tool \texttt{PerforAD}, which we release with this work.\footnote{PerforAD v1.1, \texttt{https://github.com/jhueckelheim/PerforAD/releases/tag/1.1}}

Before describing the transformations for arbitrary loop nest depths and stencil shapes, we begin with a brief summary of AD, followed by a simple example to illustrate adjoint stencils.

\subsection{Automatic Differentiation}
AD computes partial derivatives of each individual statement in a given \emph{primal} computer program and implements a new \emph{adjoint} program that accumulates those partial derivatives, following the chain rule, to obtain the derivative of the entire primal program. For example, if the original program contains a statement such as
\begin{lstlisting}
  r = sin(u);
\end{lstlisting}
then the adjoint program will contain the corresponding statement
\begin{lstlisting}
  ub = cos(u)*rb;
\end{lstlisting}
where \texttt{ub} is used to store the derivative of the program output with respect to \texttt{u} and \texttt{rb} holds the derivative of the program output with respect to \texttt{r}. The multiplication of the derivative of this statement (\texttt{cos(u)}) with the output adjoint \texttt{rb} is an application of the chain rule of calculus.
If a primal expression contains nonlinear functions (in this case \texttt{sin()}), the computation of the partial derivatives of that expression needs to access the original primal value (in this case, \texttt{u}).

An AD tool often distinguishes between active and passive variables. For example, a user may be interested only in the derivatives of some outputs with respect to some inputs. The inputs and outputs of interest are called independent and dependent, respectively. All intermediate variables that depend on an independent input and that influence a dependent output are called active; all other variables are called passive. Only active variables require a derivative counterpart variable, and only expressions that involve an active variable need to be differentiated. \texttt{PerforAD} supports these requirements by allowing a user to specify which arrays are active.

\texttt{PerforAD} is not a general-purpose AD tool. Rather, it can be seen as an AD-aware loop transformation tool that focuses on the efficient creation of adjoint stencil loops. A general-purpose AD tool is currently necessary to differentiate the entire program, except for the stencil loops that can be handled by \texttt{PerforAD}. The tool does not contain its own parser front-end and instead relies on the caller to supply a high-level description of the stencil computation. In the test cases in our paper, this description was manually created. Automating this process remains future work. \texttt{PerforAD} is designed in a modular fashion to simplify the creation of new front-ends (for example, to parse Fortran or C code) and back-ends (for example, to print Fortran or C code). \texttt{PerforAD} is written in Python and makes extensive use of the symbolic math library SymPy~\cite{10.7717/peerj-cs.103} for its internal computation, as well as for its user interface.

\subsection{1D Example}
\label{sec:1dex}
\begin{figure}
    \centering
    \def\svgwidth{1.0\linewidth}
    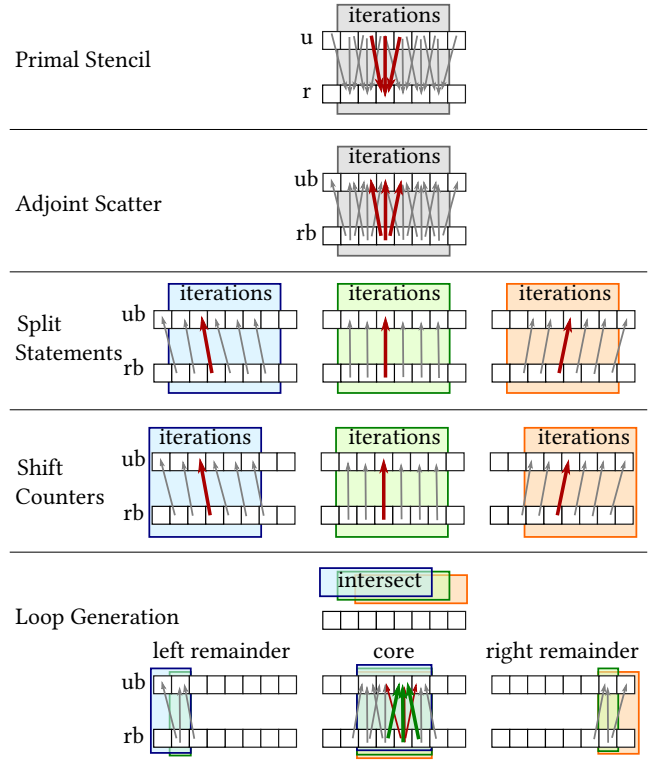
    \caption{Adjoint stencil transformation step by step, for a one-dimensional three-point stencil. This illustration does not show the read accesses to the primal array \texttt{u} that may still be needed in the derivative code. Red arrows in the primal and adjoint show a group of updates that is performed in the same loop iteration. Those same updates are then highlighted in the split, shifted, and core loop. After regeneration, the red updates are distributed among several iterations, and one iteration in the core loop performs a set of updates as shown by green arrows.}
    \label{fig:1dexsample}
\end{figure}

This section explains the adjoint stencil transformation step by step using a one-dimensional example. This explanation is accompanied by the illustration in Figure~\ref{fig:1dexsample}.

Suppose that a primal program contains the following parallel gather operation for an iteration space $i\in[1,n-1]$:
\begin{lstlisting}
#pragma omp parallel for private(i)
for ( i=1; i<=n - 1; i++ ) {
  r[i] = c[i]*(2.0*u[i-1]-3.0*u[i]+4*u[i+1]);
}
\end{lstlisting}
A straightforward reverse-mode differentiation of this loop to compute the derivatives of \texttt{r} with respect to \texttt{u} would yield the following scatter operation, where \texttt{ub} and \texttt{rb} represent the adjoint variables that correspond to \texttt{u} and \texttt{r}:
\begin{lstlisting}
for ( i=1; i<=n-1; i++ ) {
  ub[i-1] += 2.0 * c[i] * rb[i];
  ub[i] -= 3.0 * c[i] * rb[i];
  ub[i+1] += 4.0 * c[i] * rb[i];
}
\end{lstlisting}
If we assume that floating-point summation is associative (we come back to this point in Section~\ref{sec:assoc}), we can split this into three loops.
\begin{lstlisting}
for ( i=1; i<=n-1; i++ ) {
  ub[i-1] += 2.0 * c[i] * rb[i];
}
for ( i=1; i<=n-1; i++ ) {
  ub[i] -= 3.0 * c[i] * rb[i];
}
for ( i=1; i<=n-1; i++ ) {
  ub[i+1] += 4.0* c[i] * rb[i];
}
\end{lstlisting}
We can now substitute the loop counter \texttt{i} in the first, second, and third loop with \texttt{j:=i-1}, \texttt{j:=i}, and \texttt{j:=i+1}, respectively, and obtain three loops that each use the loop counter $j$ as write index and have iteration spaces $j\in[0,n-2]$, $j\in[1,n-1]$ and $j\in[2,n]$.
\begin{lstlisting}
for ( j=0; j<=n-2; j++ ) {
  ub[j] += 2.0 * c[j+1] * rb[j+1];
}
for ( j=1; j<=n-1; j++ ) {
  ub[j] -= 3.0 * c[j] * rb[j];
}
for ( j=2; j<=n; j++ ) {
  ub[j] += 4.0 * c[j-1] * rb[j-1];
}
\end{lstlisting}
We observe that the iteration space of the three loops intersects for $j\in[2,n-2]$. Only the iteration space of the first loop contains $j=0$, that of the first and second loops contain $j=1$, that of the second and third contain $j=n-1$, and the iteration space of the third loop contains $j=n$. We can therefore compute the same result using the following parallel loop and remainder statements.
\begin{lstlisting}
ub[0] += 2.0 * c[1] * rb[1];
ub[1] += 2.0 * c[2] * rb[2];
ub[1] -= 3.0 * c[1] * rb[1];
ub[n-1] -= 3.0 * c[n-1] * rb[n-1];
ub[n-1] += 4.0 * c[n-2] * rb[n-2];
ub[n] += 4.0 * c[n-1] * rb[n-1];
#pragma omp parallel for private(j)
for ( j=2; j<=n-2; j++ ) {
  ub[j] += 2.0 * c[j+1] * rb[j+1];
  ub[j] -= 3.0 * c[j] * rb[j];
  ub[j] += 4.0 * c[j-1] * rb[j-1];
}
\end{lstlisting}
Assuming that $n$ is sufficiently large, the time spent executing the remainder statements will be insignificant compared with that spent inside the loop, which contains only updates to \texttt{ub[j]} that can easily be merged into a single statement to obtain
\begin{lstlisting}
#pragma omp parallel for private(j) shared(rb,ub,c)
for ( j=2; j<=n-2; j++ ) {
  ub[j] += 4.0 * c[j-1] * rb[j-1]
         - 3.0*c[j] * rb[j]
         + 2.0 * c[j+1] * rb[j+1];
}
\end{lstlisting}
This adjoint stencil loop has the same set of read and write indices and can be parallelised in the same way as the primal stencil loop. Note that the constant factors $4.0$ and $2.0$ have swapped their position compared with the primal stencil.

\subsection{Multidimensional Stencils}
We now present the generation of adjoint stencil loops for any number of dimensions and for any stencil shape, as implemented in \texttt{PerforAD}.

\subsubsection{From Inputs to Adjoint Statements}
The tool requires as input a symbolic expression that represents the computation performed by a single iteration of the innermost loop. In addition, a list of symbolic objects representing loop counters and symbolic expressions for loop bounds is required. Also, the user must provide a list of active variables for which counterpart derivative variables will be generated. Two complete examples of \texttt{PerforAD} input scripts are given in Section~\ref{sec:testcases}.

If the loop body is sufficiently simple, the provided expression is differentiated with respect to all active input variables separately using SymPy's symbolic differentiation capabilities. We note that even though we use symbolic differentiation for the statement inside the innermost loop, the overall derivative is assembled by using automatic differentiation techniques.
For large loop bodies where symbolic differentiation is too inefficient, instead of providing a concrete expression such as $$r_{i,j} = u_{i-1,j}+2u_{i,j-1},$$ the user can provide an uninterpreted function such as $$r_{i,j} = f(a,b)(a=u_{i-1,j}, b=u_{i,j-1}).$$ \texttt{PerforAD} will then generate adjoint code where the derivatives of $f$ are also uninterpreted function calls. For example, the derivative of $f$ with respect to its input $a$, evaluated at $(u_{i-1,j},u_{i,j-1})$, would be written as $$r_{i,j} = \textit{derivative}(f(a,b),a)(a=u_{i-1,j}, b=u_{i,j-1}).$$ A back-end could easily replace this expression by the appropriate call to a derivative function that was created manually or by AD.

Each of these generated expressions represents the partial derivative of the stencil with respect to one input (for example, $\frac{\partial f}{\partial L}$). The derivative expressions that are generated in this way will be denoted by $S_1, S_2, \ldots, S_M$, where the number of such expressions $M$ is  bounded by the stencil size. For example, the above two-point stencil would yield $M=2$. In order to implement back-propagation, each expression needs to be multiplied with the adjoint variable that is associated with the output of the primal loop body (in this example, $\bar{r}_{i,j}$), and the result must be assigned to the adjoint variable that is associated with the input with respect to which this expression was differentiated (in this example, $\bar{u}_{i-1,j}$). This yields the expression $\bar{u}_{i-1,j}\; += \frac{\partial f}{\partial a}\cdot \bar{r}_{i,j}$. Without the following steps, the derivative expressions would collectively implement the scatter operation that is typical for conventional adjoints of stencil loops.

\subsubsection{Shifting the Indices}
To transform the scatter into a gather operation, \texttt{PerforAD} determines the constant offset $o$ that is used by each derivative expression (in the above example, a two-dimensional vector $o=[-1;0]$). All indices of that expression are increased by $-o$. In our example, this yields $\bar{u}_{i,j} += \frac{\partial f}{\partial L}\cdot \bar{r}_{i+1,j}$. In order to preserve the semantics of the adjoint computation, the offset is stored along with the shifted expression and is later used to adjust the loop bounds accordingly. The result of this step is a list of tuples, each containing an expressions and an offset. All expressions now write to the same output index.

If the output of the primal stencil function depends on its inputs in a nonlinear way, then the derivative computation requires read access to the primal input, and any array accesses also need to occur with shifted indices. For our example with $$\textit{derivative}(f(a,b),a)(a=u_{i-1,j}, b=u_{i,j-1}),$$ the derivative with offsets shifted by $o=[-1;0]$ would become $$\textit{derivative}(f(a,b),a)(a=u_{i,j}, b=u_{i+1,j-1}).$$
In some cases, the shifted derivative functions will read data from indices that did not occur in any primal expression (here, $i+1,j-1$).

\subsubsection{Generating the Core Loop Nest}
The offsets are now used to identify the largest iteration space in which it is legal to execute all shifted derivative expressions generated in the previous step. We will refer to the loop that covers this space as the \emph{core loop nest}. The loop bounds of the core loop nest depend on the bounds of the primal loop nest supplied by the user, as well as the offsets of the shifted expressions. In each dimension $i$, the iteration space of the primal loop will be denoted as $i\in[s_i, e_i]$ where $s_i, e_i$ refer to the primal lower and primal upper bound, respectively. The $i$ component of the offset vector of a given adjoint expression $S_l$will be denoted as $o_i(S_l)$. The lower core loop bound is then given by the original lower loop bound given by the user, plus the maximum offset in that dimension. Conversely, the upper loop bound is given by the original upper loop bound minus the minimum offset. Formally, the adjoint core loop bounds in $i$-dimension are $$\left[s_i+\max\limits_{l \in [1,M]}(o_i(S_l)),\quad s_e-\min\limits_{l \in [1,M]}(o_i(S_l))\right].$$

The core loop nest implements a gather operation with as many distinct read indices as the primal loop. If the iteration space is much larger than the stencil size, this loop nest will perform most of the adjoint computation. The core loop nest can be easily parallelised.

\subsubsection{Boundary Treatment}
\begin{figure}
\includegraphics[width=0.6\linewidth]{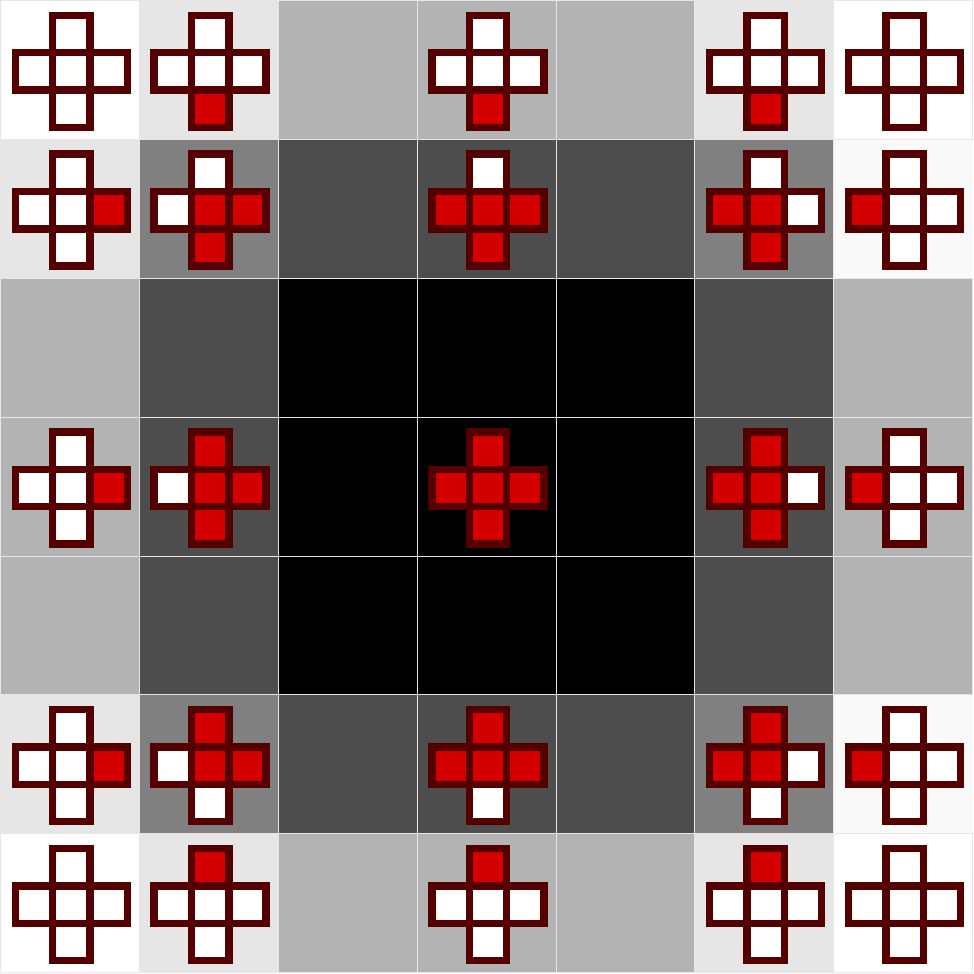}
\caption{Iteration spaces of adjoint stencil loop nests for a five-point stencil in two dimensions (that is, reading from $[i-1,j],[i+1,j],[i,j],[i,j-1],[i,j+1]$). Each of the large squares represents one point in the iteration space. Areas with uniform colour are part of the same loop nest. The core loop is represented by the dark area in the centre of the image. Within each loop nest, a stencil gather operation is implemented. The read indices are a subset of the original five-point star stencil. The subset that is actually implemented in that particular loop nest is shown by filled red boxes; the expressions that are not valid in this area are shown with empty red boxes.}
\label{fig:splitting}
\end{figure}
In addition to the core loop nest, we need to generate boundary loop nests that execute the appropriate subset of derivative expressions wherever they are valid outside the core area. To achieve these, we split the set of derivative expressions into subsets such that each subset contains only expressions with the same iteration space. Usually more than one correct splitting exists. In the current implementation, the loop nests are generated such that their iteration spaces are disjoint; that is, every output index is accessed only within one loop nest that contains all statements that update this location. As a result, no synchronisation barriers are needed between the generated loop nests. Note that the boundary loops are also stencil gather operations, albeit with smaller numbers of read indices, and can be easily parallelised; see Figure~\ref{fig:splitting}.

The current splitting strategy results in a relatively large code size: with a primal stencil that gathers data from $n$ points in each of $d$ dimensions, the number of generated adjoint loop nests is at most $(2n-1)^d$. For the one-dimensional three-point stencil in Section~\ref{sec:1dex}, this resulted in five adjoint loops, of which all but the core loop could be unrolled, since they performed only one iteration. For a dense $3\times3$ stencil in two dimensions, the number of adjoint loops would be $25$, and for a dense three-dimensional $3\times3\times3$ stencil, $125$. If the primal stencil is not dense but instead a star-shaped stencil such as the one shown in Section~\ref{sec:testcases3dwave}, then $53$ loop nests are needed.

To avoid generating such a large amount of code, the strategy could be changed to instead generate only one remainder loop on each side of the core loop in each dimension, containing all derivative expressions. Since not all expressions are valid throughout the whole remainder, each statement would need to be guarded by an \texttt{if}-condition. While this might not be ideal for performance, the effect would be limited to the remainder loop nests, which are at most $d-1$-dimensional. A polyhedral compiler could be used to implement many other strategies, with different trade-offs between the code size and the number of branches.

Another strategy is possible if the primal stencil is always executed on entire arrays and the AD tool can control the memory allocation for both primal and adjoint code. In this case, the input arrays could be padded with zeroes, which would allow all shifted adjoint expressions to be executed throughout the entire domain.

\subsection{Restrictions}
The transformation implemented in \texttt{PerforAD} requires that there be a set of input arrays from which data is read and a set of output arrays that are written in an assignment or incremented by using the \texttt{+=} operator. The sets of read and write arrays must not intersect; that is, no array can be both input and output (with the aforementioned exception of the \texttt{+=} operator). In addition, read access is allowed from constants that are not active for differentiation. All output arrays are accessed only by using the loop counters as indices. For a nested loop with counter variables $[i_1,i_2,\ldots]$, a multidimensional array may be written or incremented by using a permuted subset of the counter variables as indices, for example, \texttt{r[i\_1][i\_4][i\_3]}. All input arrays are read at indices that are constant integer offsets of the loop counters, for example, \texttt{u[i\_1+3][i\_3-2][i\_4-1]}. The offsets must be known at compile time for expressions to be shifted and remainder loops to be generated.
The loop nest must be perfect; that is, no statement may appear other than inside the innermost loop. All loop bounds must be affine functions.

\subsection{Associativity of Floating-Point Summation}
\label{sec:assoc}
We assume that the contributions to each adjoint output index can be reordered arbitrarily without affecting the final result. This assumption would be true for real numbers but is not true for floating-point numbers because of roundoff effects. If this effect is important for some application, \texttt{PerforAD} could be modified to respect a particular ordering that the user chooses. The ordering would still be deterministic if the adjoint code is executed in parallel. This is possible because all updates to a given index are collected within the same iteration. In contrast, the scatter operation generated by a conventional AD tool could not be easily parallelised in a way that produces a deterministic floating-point result.

\subsection{Verification}
To verify the correctness of the adjoint code generated by \texttt{PerforAD}, we compared it 
with adjoint code generated by the AD tools \texttt{ADIC}~\cite{narayanan2010adic2} and 
\texttt{Tapenade}~\cite{tapenade}. We used the test cases described in Section~\ref{sec:testcases} and checked 
for equality of the computed adjoints. For all the input test cases in this paper, the outputs of all 
the three approaches were in full agreement.

\section{Test Cases}
\label{sec:testcases}

We have selected two test cases to demonstrate the performance of adjoint stencils in \texttt{PerforAD}. Both test cases are structured-mesh discretisations of partial differential equations. The generated adjoint stencil code faces two performance challenges. First, a large number of remainder loops may cause large adjoint code sizes and potentially slow the execution. Second, the use of symbolic differentiation applied to the loop body may cause unnecessary computations for complicated primal loop bodies, especially since \texttt{PerforAD} makes no attempt to identify common sub-expressions  within the same loop nest. We selected a test case with a deep loop nest and another one with a complicated loop body to test our tool in the presence of these challenges.

The first test case is a solver for wave equations  with spatially varying wave propagation speed. The computation is performed in a three-deep loop nest, corresponding to the three spatial dimensions. A large number of iterations is typically performed with this type of solver, which makes this a good showcase for performance evaluations.

The second test case is a one-dimensional Burgers equation solver. The loop body is more challenging to differentiate, because it is nonlinear and only piecewise differentiable. As a result, the generated adjoint is more complex and contains branches.

For both test cases, we generate adjoint stencils using \texttt{PerforAD} and generate conventional adjoint code with scatter memory access using the Tapenade~\cite{tapenade} AD tool. Tapenade  generates serial output code. We thus additionally create a parallelised version of the conventional adjoint code by adding OpenMP pragmas to the Tapenade-generated code, safeguarding the scattered memory access with \texttt{atomic} pragmas to avoid data races.

\subsection{Wave Equation}
\label{sec:testcases3dwave}
The wave equation and its adjoint are solved numerically in applications including seismic imaging, where performance on high-performance computing systems has been identified as a bottleneck~\cite{5628156}.
We perform one single time step on a $1000\times1000\times1000$ three-dimensional grid.
The wave equation is given by
\begin{align*}
\frac{\partial^2 u}{\partial t^2} &= a^2 \cdot 
         \left( \frac{\partial^2 u}{\partial x^2} +
        \frac{\partial^2 u}{\partial y^2} + \frac{\partial^2 u}{\partial z^2}\right) \hbox{ for }\vec{x}\in\Omega\subset\mathbb{R}^3,\ t\in (0,T].
\end{align*}
By using finite differences to discretise the differential operators in space and time with spatial and temporal resolution $dx$ and $dt$ and by letting $c:=a^2$ and $$D := \left(\frac{dt}{dx}\right)^2 = \left(\frac{dt}{dy}\right)^2 = \left(\frac{dt}{dz}\right)^2,$$ a primal solver to solve this equation, as well as its adjoint, can be generated with the Python script shown in Figure~\ref{fig:perforadwave}.

\begin{figure}
    \centering
\begin{lstlisting}[language=Python]
import sympy as sp
from perforad import *
######## 3D Wave Equation Example ########
# Define symbols
c = sp.Function("c")
u_1 = sp.Function("u_1"); u_1_b = sp.Function("u_1_b")
u_2 = sp.Function("u_2"); u_2_b = sp.Function("u_2_b")
u = sp.Function("u")    ; u_b = sp.Function("u_b")
i,j,k,C,D,n = sp.symbols("i,j,k,C,D,n")
# Build stencil expression
u_xx = u_1(i-1,j,k) - 2*u_1(i,j,k) + u_1(i+1,j,k)
u_yy = u_1(i,j-1,k) - 2*u_1(i,j,k) + u_1(i,j+1,k)
u_zz = u_1(i,j,k-1) - 2*u_1(i,j,k) + u_1(i,j,k+1)
expr = 2.0*u_1(i,j,k) - u_2(i,j,k) + c(i,j,k)*D*(u_xx+u_yy+u_zz)
# Build LoopNest object for this expression
lp = makeLoopNest(lhs=u(i,j,k), rhs=expr, counters = [i,j,k],
                  bounds={i:[1,n-2],j:[1,n-2],k:[1,n-2]})
# Output primal and adjoint files
printfunction(name="wave3d", loopnestlist=[lp])
printfunction(name="wave3d_perf_b",
              loopnestlist=lp.diff({u:u_b, u_1:u_1_b, u_2: u_2_b}))
\end{lstlisting}
    \caption{Python script that uses the \texttt{PerforAD} package to generate C code, which implements a three-dimensional wave equation solver. The input and output arrays are represented in \texttt{PerforAD} using SymPy function objects. All scalars, including loop counters and bounds, are represented by SymPy symbol objects. The SymPy expression object \texttt{expr}, a list of loop counters, and a dictionary with bounds associated with each counter are passed to the \texttt{makeLoopNest()} function, which is part of \texttt{PerforAD}. It returns a \texttt{LoopNest} object, which encapsulates all necessary data to represent a stencil computation, and provides functions to generate the primal and adjoint C code.}
    \label{fig:perforadwave}
\end{figure}

The primal and adjoint stencil loop shown in Figure~\ref{fig:primalwave} are generated by \texttt{PerforAD} and shown alongside the conventional adjoint generated by Tapenade, which is not a stencil loop and thus had to be manually parallelised.

\begin{figure}
    \centering
\begin{lstlisting}
// Stencil primal
#pragma omp parallel for private(k,j,i)
for ( i=1; i<=n - 2; i++ )
  for ( j=1; j<=n - 2; j++ )
    for ( k=1; k<=n - 2; k++ ) {
      u[i][j][k] += D*(-6*u_1[i][ j][ k] + u_1[i][ j][ k - 1]
                     + u_1[i][ j][ k + 1] + u_1[i][ j - 1][ k]
                     + u_1[i][ j + 1][ k] + u_1[i - 1][ j][ k]
                     + u_1[i + 1][ j][ k])*c[i][ j][ k]
                     + 2.0*u_1[i][ j][ k] - u_2[i][ j][ k];
}
\end{lstlisting}
\begin{lstlisting}
// PerforAD stencil adjoint
for ( i=2; i<=n - 3; i++ )
  for ( j=2; j<=n - 3; j++ )
    for ( k=2; k<=n - 3; k++ ) {
      u_1_b[i][j][k] += D*c[i][ j][ k + 1]*u_b[i][ j][ k + 1];
      u_1_b[i][j][k] += D*c[i][ j + 1][ k]*u_b[i][ j + 1][ k];
      u_1_b[i][j][k] += D*c[i + 1][ j][ k]*u_b[i + 1][ j][ k];
      u_1_b[i][j][k] += (-6*D*c[i][ j][ k] + 2.0)*u_b[i][ j][ k];
      u_2_b[i][j][k] += -u_b[i][ j][ k];
      u_1_b[i][j][k] += D*c[i - 1][ j][ k]*u_b[i - 1][ j][ k];
      u_1_b[i][j][k] += D*c[i][ j - 1][ k]*u_b[i][ j - 1][ k];
      u_1_b[i][j][k] += D*c[i][ j][ k - 1]*u_b[i][ j][ k - 1];
}
\end{lstlisting}
\begin{lstlisting}
// Tapenade adjoint, manually parallelised
#pragma omp parallel for private(i, j, k, tempb)
for (i = n-2; i > 0; --i)
  for (j = n-2; j > 0; --j)
    for (k = n-2; k > 0; --k) {
      tempb = D*c[i][j][k]*ub[i][j][k];
      #pragma omp atomic
      u_1b[i][j][k - 1] = u_1b[i][j][k - 1] + tempb;
      #pragma omp atomic
      u_1b[i][j][k] = u_1b[i][j][k] + 2.0*ub[i][j][k] - 6*tempb;
      #pragma omp atomic
      u_1b[i][j][k + 1] = u_1b[i][j][k + 1] + tempb;
      #pragma omp atomic
      u_1b[i][j - 1][k] = u_1b[i][j - 1][k] + tempb;
      #pragma omp atomic
      u_1b[i][j + 1][k] = u_1b[i][j + 1][k] + tempb;
      #pragma omp atomic
      u_1b[i - 1][j][k] = u_1b[i - 1][j][k] + tempb;
      #pragma omp atomic
      u_1b[i + 1][j][k] = u_1b[i + 1][j][k] + tempb;
      u_2b[i][j][k] = u_2b[i][j][k] - ub[i][j][k];
}
\end{lstlisting}
    \caption{One single time step for the primal and adjoint wave equation solver implemented in C, as generated by \texttt{PerforAD}. The output array is written to at the central index \texttt{i,j,k}, while the input arrays are read in the neighbourhood around that centre. The loop is parallelised. In contrast, the conventional adjoint implements a scatter operation to a neighbourhood around the centre and was manually parallelised by using atomic updates.}
    \label{fig:primalwave}
\end{figure}

\subsection{Burgers Equation}
Computational fluid dynamics (CFD) is another application that routinely uses adjoints and relies on efficient parallelisation to compute industrial-scale test cases. The Burgers equation is often used as a prototype in CFD solver development, because it is scalar and thus relatively easy to implement, but it still shows some of the challenging nonlinear behaviour that makes CFD simulations difficult. In one spatial dimension, it can be written as
\begin{align*}
    \frac{\partial u}{\partial t} + u \frac{\partial u}{\partial x} = \nu \; \frac{\partial ^2 u}{\partial x^2}.
\end{align*}
An interesting challenge in the context of this work is that the convective term is nonlinear. A common way of discretising this term is \emph{upwinding}, where the finite difference formula depends on the sign of $u$ as
$$
u \frac{\partial u}{\partial x} \approx \frac{1}{dx}\cdot\left(\max(u_i,0)\cdot (u_i-u_{i-1})+\min(u_i,0)\cdot (u_{i+1}-u_i)\right).
$$
By letting $C := \frac{dt}{dx}$ and $D := \nu * \frac{dt}{dx^2}$, the primal and adjoint stencils can be implemented by using the \texttt{PerforAD} script shown in Figure~\ref{fig:burgersperf}, yielding the C code shown in Figure~\ref{fig:burgerscodes}.

\begin{figure}
    \centering
\begin{lstlisting}[language=Python]
import sympy as sp
from perforad import *
######## 1D Burgers Equation Example ########
# Build stencil expression
ap = sp.functions.Max(u_1(i),0)
am = sp.functions.Min(u_1(i),0)
uxm = u_1(i)-u_1(i-1)
uxp = u_1(i+1)-u_1(i)
ux = ap*uxm+am*uxp
expr = u_1(i) - C * ux + D * (u_1(i+1) + u_1(i-1) - 2.0*u_1(i))
# Build LoopNest object for this expression
lp = makeLoopNest(lhs=u(i), rhs=expr, counters = [i],
                  bounds={i:[1,n-2]})
# Output primal and adjoint files
printfunction(name="burgers1d", loopnestlist=[lp])
printfunction(name="burgers1d_perf_b",
              loopnestlist=lp.diff({u:u_b, u_1:u_1_b}))
\end{lstlisting}
    \caption{Python script to generate the Burgers equation solver. An upwinding scheme is implemented by using $\min$ and $\max$ functions, which are piecewise differentiable and correctly handled by SymPy and \texttt{PerforAD}.}
    \label{fig:burgersperf}
\end{figure}

The stencil is nonlinear, and the value of $u$ is required at various points in the derivative computation.
The adjoint code generated by Tapenade handles the $\min$ and $\max$ functions differently from \texttt{PerforAD}. Instead of evaluating these functions within the adjoint code, Tapenade creates a loop that evaluates the functions separately and pushes the results onto a stack. These precomputed values are popped from the stack during the adjoint computation. The resulting code cannot easily be parallelised, since the order in which values are added and removed from the stack is crucial.

To exclude this effect and compare \texttt{PerforAD} with a parallelised adjoint code that uses scatter operations, we manually modified the Tapenade output to instead evaluate the $\min$ and $\max$ functions within the adjoint computation itself, and we removed the stack access. Following this, we made the loop OpenMP-parallel and added \texttt{atomic} pragmas. We ran the primal and all adjoint solvers on a one-dimensional problem with one time step on 1 billion cells.

\begin{figure}
    \centering
\begin{lstlisting}
// Stencil Primal
for ( i=1; i<=n - 2; i++ ) {
    u[i] += -C*((-u_1[i] + u_1[i + 1])*fmin(0, u_1[i])
               + (u_1[i] - u_1[i - 1])*fmax(0, u_1[i]))
           + D*(-2.0*u_1[i] + u_1[i - 1] + u_1[i + 1])
           + u_1[i];
}
\end{lstlisting}
\begin{lstlisting}
// PerforAD Stencil Adjoint
for ( i=2; i<=n - 3; i++ ) {
    u_1_b[i] += (C*fmax(0, u_1[i + 1]) + D)*u_b[i + 1];
    u_1_b[i] += (-C*((-u_1[i] + u_1[i + 1])*((-u_1[i]>=0)?1.0:0.0)
                    + (u_1[i] - u_1[i - 1])*((u_1[i]>=0)?1.0:0.0)
                    + fmax(0, u_1[i]) - fmin(0, u_1[i]))
                - 2.0*D + 1)*u_b[i];
    u_1_b[i] += (-C*fmin(0, u_1[i - 1]) + D)*u_b[i - 1];
}
\end{lstlisting}
    \caption{Primal and adjoint stencil solvers for the Burgers equation in one dimension. The upwinding scheme results in the use of ternary operators in the adjoint code.}
    \label{fig:burgerscodes}
\end{figure}

\section{Experimental Results}
\label{sec:results}

\pgfplotstableread{
threads priwave   adjwave   perfwave   atomicwave priwaveS  adjwaveS  perfwaveS  atomicwaveS priburger adjburger perfburger atomicburger  priburgerS adjburgerS perfburgerS atomicburgerS
1       12.817169 25.450112 41.271231  551.062598 12.817169 25.450112 41.271231  551.062598  25.022697 95.739452 51.853798  700.108687    25.022697  95.739452  51.853798   700.108687
2       6.497922  25.517687 21.191146  297.446950 12.817169 25.450112 41.271231  551.062598  12.811893 96.697623 26.264378  682.156979    25.022697  95.739452  51.853798   700.108687
4       3.303470  25.493159 10.685106  573.912748 12.817169 25.450112 41.271231  551.062598  6.415169  97.710026 13.046761  1584.486628   25.022697  95.739452  51.853798   700.108687
8       1.708070  25.501117 5.403113   634.263056 12.817169 25.450112 41.271231  551.062598  3.234178  96.527791 6.509971   1836.238670   25.022697  95.739452  51.853798   700.108687
16      0.917104  25.507034 2.716764   667.388490 12.817169 25.450112 41.271231  551.062598  1.656313  96.200815 3.257029   nan           25.022697  95.739452  51.853798   700.108687
32      0.843320  25.488998 1.437381   nan        12.817169 25.450112 41.271231  551.062598  0.858910  95.280220 1.635134   nan           25.022697  95.739452  51.853798   700.108687
64      0.876202  25.502668 1.286821   nan        12.817169 25.450112 41.271231  551.062598  0.517281  94.502392 0.871684   nan           25.022697  95.739452  51.853798   700.108687
128     nan       nan       nan        nan        nan       nan       nan        nan         0.502064  93.675209 0.756979   nan           25.022697  95.739452  51.853798   700.108687
256     nan       nan       nan        nan        nan       nan       nan        nan         0.524638  94.563273 0.766647   nan           25.022697  95.739452  51.853798   700.108687
}\MICTable
\pgfplotstableread{
threads  priwave   atomicwave perfwave adjwave   priwaveS  atomicwaveS perfwaveS adjwaveS  priburger atomicburger  perfburger adjburger priburgerS atomicburgerS perfburgerS adjburgerS
1        7.778035  78.799311  8.744958 4.888895  7.778035  78.799311   8.744958  4.888895  4.629663  98.837754     9.831660   5.491687  4.629663   98.837754     9.831660    5.491687
2        3.554153  165.221611 4.840905 4.888895  7.778035  78.799311   8.744958  4.888895  3.620121  892.074053    5.197782   5.491687  4.629663   98.837754     9.831660    5.491687
4        1.803566  415.370337 2.537966 4.888895  7.778035  78.799311   8.744958  4.888895  1.649297  nan           3.046296   5.491687  4.629663   98.837754     9.831660    5.491687
8        1.235498  382.043754 1.459115 4.888895  7.778035  78.799311   8.744958  4.888895  0.816954  nan           1.553153   5.491687  4.629663   98.837754     9.831660    5.491687
16       1.130346  384.424152 1.186694 4.888895  7.778035  78.799311   8.744958  4.888895  0.586842  nan           0.836597   5.491687  4.629663   98.837754     9.831660    5.491687
32       1.142283  414.113730 1.328224 4.888895  7.778035  78.799311   8.744958  4.888895  0.542569  nan           0.635112   5.491687  4.629663   98.837754     9.831660    5.491687
}\CPUTTable

\pgfplotstableread{
threads  priwave   atomicwave perfwave adjwave   priwaveS  atomicwaveS perfwaveS adjwaveS  priburger atomicburger  perfburger adjburger priburgerS atomicburgerS perfburgerS adjburgerS
1        4.135591  91.770882  8.521384 5.434366  4.135591  91.770882   8.521384  5.434366  2.129241  173.550478    15.727860  8.756135  2.129241   173.550478     15.727860    8.756135
2        2.253192  116.156586 4.331313 5.434366  4.135591  91.770882   8.521384  5.434366  1.149422  680.058104    8.036074   8.756135  2.129241   173.550478     15.727860    8.756135
4        1.278260  194.693396 2.438649 5.434366  4.135591  91.770882   8.521384  5.434366  0.706607  855.134868    4.470993   8.756135  2.129241   173.550478     15.727860    8.756135
6        1.028625  203.924252 1.718890 5.434366  4.135591  91.770882   8.521384  5.434366  0.604912  843.022027    3.048968   8.756135  2.129241   173.550478     15.727860    8.756135
8        0.923506  204.367645 1.400985 5.434366  4.135591  91.770882   8.521384  5.434366  0.572293  886.445933    2.290882   8.756135  2.129241   173.550478     15.727860    8.756135
12       0.897188  212.272802 1.322396 5.434366  4.135591  91.770882   8.521384  5.434366  0.555659  894.896142    1.535777   8.756135  2.129241   173.550478     15.727860    8.756135
16       0.993185  710.124315 1.613896 5.434366  4.135591  91.770882   8.521384  5.434366  0.587833  2991.078209   2.287722   8.756135  2.129241   173.550478     15.727860    8.756135
}\CPUTable

We obtained timing results on two test systems for the primal and adjoint, as well as conventional adjoint solvers for the wave equation, and the Burgers equation. The first test system, referred to as \emph{Broadwell}, uses a dual socket system with 12-core Intel Xeon E5-2650 v4 Broadwell CPUs, with a combined total of 24 physical cores. To eliminate NUMA effects on our experiments, we limited the number of threads to 12, and used \texttt{numactl} to restrict memory and computation to a single socket and \texttt{OMP\_PROC\_BIND} to ensure thread pinning. The other test system, referred to as \emph{KNL}, runs an Intel XeonPhi Knights Landing 7210 processor with 64 physical cores and up to 256 threads, in \texttt{KMP\_AFFINITY=scatter} mode. The examples are compiled using the Intel C Compiler version 18 using the flags \texttt{-O3 -fopenmp -xHost}.

\subsection{Broadwell}
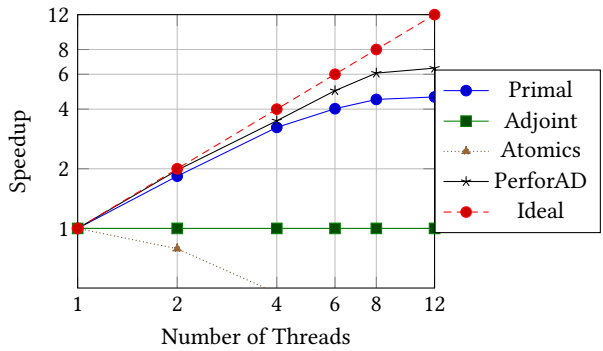
\begin{figure}
\begin{tikzpicture}
\begin{axis}[
ymin=0.5,
ymax=12,
xmin=1,
xmax=12,
legend style={at={(1,0.5)},
      anchor=west,legend columns=1},
xtick={1,2, 4,6,8, 12},
xticklabels={1,2, 4,6,8, 12},
ytick={1,2, 4,6,8, 12},
yticklabels={1,2, 4,6,8, 12},
xlabel={Number of Threads},
ylabel={Speedup},
title={Scalability of the Wave Equation on Broadwell},
title style={font=\bfseries},
xmode=log,
ymode=log,
xmajorgrids,
ymajorgrids,
width=15em,
height=11.5em,
cycle list name=cbw,
scale only axis
]
\addplot table [y expr=\thisrow{priwaveS} / \thisrow{priwave}] {\CPUTable}; \addlegendentry{Primal}
\addplot table [y expr=\thisrow{adjwaveS} / \thisrow{adjwave}] {\CPUTable}; \addlegendentry{Adjoint}
\addplot table [y expr=\thisrow{atomicwaveS} / \thisrow{atomicwave}] {\CPUTable}; \addlegendentry{Atomics}
\addplot table [y expr=\thisrow{perfwaveS} / \thisrow{perfwave}] {\CPUTable}; \addlegendentry{PerforAD}
\addplot table [y expr=\thisrow{threads}] {\CPUTable}; \addlegendentry{Ideal}
\end{axis}
\end{tikzpicture}
\caption{Speedups for the wave equation solver on a Broadwell processor, using up to 12 threads. The Tapenade-generated code with manual parallelisation does not scale at all. The primal and PerforAD-generated adjoint benefit from using all 12 cores.}
\label{fig:scalewavebwd}
\end{figure}

\begin{figure}
\begin{tikzpicture}
\begin{axis}[
ymin=0.5,
ymax=12,
xmin=1,
xmax=12,
legend style={at={(1,0.5)},
      anchor=west,legend columns=1},
xtick={1,2, 4,6,8, 12},
xticklabels={1,2, 4,6,8, 12},
ytick={1,2, 4,6,8, 12},
yticklabels={1,2, 4,6,8, 12},
xlabel={Number of Threads},
ylabel={Speedup},
title={Scalability of the Burgers Equation on Broadwell.},
title style={font=\bfseries},
xmode=log,
ymode=log,
xmajorgrids,
ymajorgrids,
width=15em,
height=11.5em,
cycle list name=cbw,
scale only axis
]
\addplot table [y expr=\thisrow{priburgerS} / \thisrow{priburger}] {\CPUTable}; \addlegendentry{Primal}
\addplot table [y expr=\thisrow{adjburgerS} / \thisrow{adjburger}] {\CPUTable}; \addlegendentry{Adjoint}
\addplot table [y expr=\thisrow{atomicburgerS} / \thisrow{atomicburger}] {\CPUTable}; \addlegendentry{Atomics}
\addplot table [y expr=\thisrow{perfburgerS} / \thisrow{perfburger}] {\CPUTable}; \addlegendentry{PerforAD}
\addplot table [y expr=\thisrow{threads}] {\CPUTable}; \addlegendentry{Ideal}
\end{axis}
\end{tikzpicture}
\caption{Speedups for the Burgers equation solver. The PerforAD-generated adjoint has near-perfect scalability.}
\label{fig:scaleburgerbwd}
\end{figure}
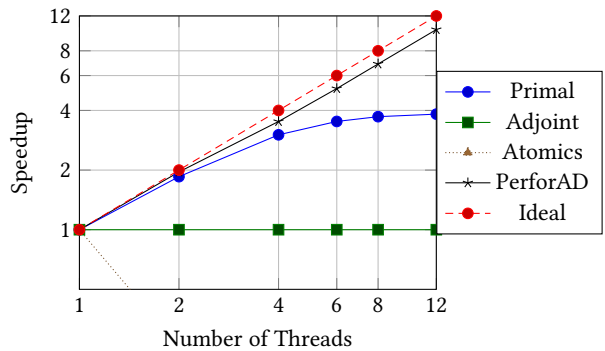

\begin{figure}
\begin{tikzpicture}
  \begin{axis}[
    ybar,
    ylabel={Runtime (s)},
    legend style={at={(0,1)},
    anchor=north west,legend columns=1},
    xtick={0,1,2,3,4},
    xticklabels={Primal Serial, PerforAD Serial, Adjoint Serial, Primal Parallel, PerforAD Parallel },
    xtick=data,
    title={Runtimes of the Wave Equation on Broadwell},
    title style={font=\bfseries},
    ymajorgrids,
    nodes near coords,
	nodes near coords align={above},
    point meta=rawy,
    x tick label style={rotate=20,anchor=east},
    width=15em,
    height=11.5em,
	scale only axis,
	ymin=0.0,ymax=10,
	xmin=-0.5,xmax=4.5,
	clip=false
    ]
    \addplot[fill=black!70,draw=none] coordinates {
		(0, 4.135591)
		(1, 8.521384)
		(2, 5.434366)
       (3, 0.897188)
		(4, 1.613896)
		};

\end{axis}
\end{tikzpicture}
\caption{Absolute runtimes for wave equation primal and adjoint stencils and conventional adjoints in serial, as well as best observed primal and adjoint stencil run time in parallel. The best-observed performance of adjoint stencils was with 12 threads and is faster than the conventional adjoint by a factor of $3.4\times$.}
\label{fig:timewavebwd}
\end{figure}
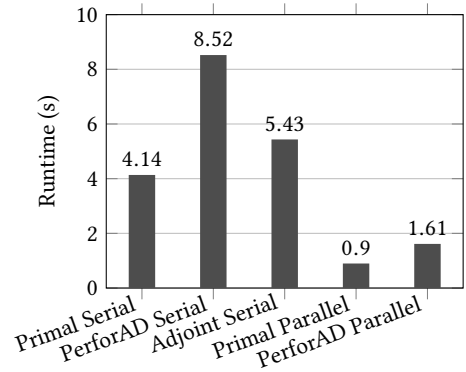

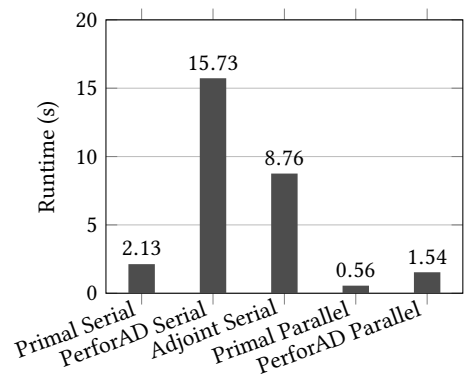
\begin{figure}
\begin{tikzpicture}
  \begin{axis}[
    ybar,
    ylabel={Runtime (s)},
    legend style={at={(0,1)},
      anchor=north west,legend columns=1},
    xtick={0,1,2,3,4}, xticklabels={Primal Serial, PerforAD Serial, Adjoint Serial, Primal Parallel, PerforAD Parallel },
    xtick=data,
    title={Runtimes of the Burgers Equation on Broadwell},
title style={font=\bfseries},
ymajorgrids,
    nodes near coords,
	nodes near coords align={above},
point meta=rawy,
    x tick label style={rotate=20,anchor=east},
width=15em,
height=11.5em,
	scale only axis,
	ymin=0.0,ymax=20,
	xmin=-0.5,xmax=4.5,
	clip=false
    ]
    \addplot[fill=black!70,draw=none] coordinates {
		(0, 2.129241)
		(1, 15.727860)
		(2, 8.756135)
        (3, 0.555659)
		(4, 1.535777)
		};

\end{axis}
\end{tikzpicture}
\caption{Absolute runtimes for the Burgers equation primal and adjoint stencils and conventional adjoints in serial, as well as best observed primal and adjoint stencil runtime in parallel. Despite being slower in serial, the adjoint stencil outperforms conventional adjoints by a factor of $5.7\times$.}
\label{fig:timeburgerbwd}
\end{figure}

Since \texttt{PerforAD} introduces overhead by differentiating the loop body separately for each function input using SymPy, we expect the serial performance of the generated adjoint stencils to be worse than that of Tapenade-generated adjoint code. On the other hand, we expect adjoint stencils to have parallel scalability at least equivalent to that of the primal stencil.

All of this is indeed confirmed in our experiments. Figure~\ref{fig:timewavebwd} shows absolute runtimes for our wave equation test case on Broadwell. The serial runtime of adjoint stencils is $64\%$ higher than that of Tapenade adjoints, at $8.52 s$ compared with $5.43 s$. However, using \texttt{atomic} pragmas on the conventional adjoint slows that code, and it requires $91 s$ even if only one thread is used, and it slows down further with any additional thread that is added to the computation. This result is in agreement with the reported slowdown of using atomics in~\cite{ragan2013halide}. In contrast, the adjoint stencil code scales well and performs slightly better than the conventional adjoint if $2$ threads are used, and significantly better if more threads are added.
Similarly, Figure~\ref{fig:timeburgerbwd} shows absolute runtimes for the Burgers equation test case, where the serial performance of adjoint stencils is worse but, because of the improved scalability, outperforms the conventional adjoint solver if at least $2$ threads are used.
The scalability of all wave equation solvers on Broadwell is presented in Figure~\ref{fig:scalewavebwd}, and for the Burgers equation solvers in Figure~\ref{fig:scaleburgerbwd}.

\subsection{Knights Landing (KNL)}

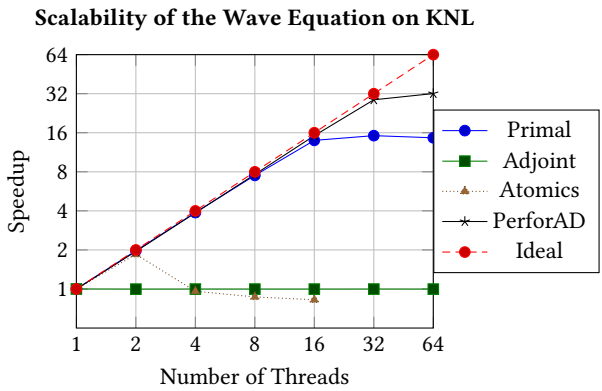
\begin{figure}
\begin{tikzpicture}
\begin{axis}[
ymin=0.5,
ymax=64,
xmin=1,
xmax=64,
legend style={at={(1,0.5)},
      anchor=west,legend columns=1},
xtick={1,2, 4,8, 16, 32, 64},
xticklabels={1,2, 4,8, 16, 32, 64},
ytick={1,2, 4,8, 16, 32, 64},
yticklabels={1,2, 4,8, 16, 32, 64},
xlabel={Number of Threads},
ylabel={Speedup},
title={Scalability of the Wave Equation on KNL},
title style={font=\bfseries},
xmode=log,
ymode=log,
xmajorgrids,
ymajorgrids,
width=15em,
height=11.5em,
cycle list name=cbw,
scale only axis
]
\addplot table [y expr=\thisrow{priwaveS} / \thisrow{priwave}] {\MICTable}; \addlegendentry{Primal}
\addplot table [y expr=\thisrow{adjwaveS} / \thisrow{adjwave}] {\MICTable}; \addlegendentry{Adjoint}
\addplot table [y expr=\thisrow{atomicwaveS} / \thisrow{atomicwave}] {\MICTable}; \addlegendentry{Atomics}
\addplot table [y expr=\thisrow{perfwaveS} / \thisrow{perfwave}] {\MICTable}; \addlegendentry{PerforAD}
\addplot table [y expr=\thisrow{threads}] {\MICTable}; \addlegendentry{Ideal}
\end{axis}
\end{tikzpicture}
\caption{Speedups for the wave equation solver on KNL, using up to 64 threads. The primal loop scales well only up to 16 threads. The adjoint stencil loop generated by \texttt{PerforAD} continues to scale up to 32 threads. The manually parallelised adjoint solver scales only up to 2 threads.}
\label{fig:scalewave}
\end{figure}

\begin{figure}
\begin{tikzpicture}
\begin{axis}[
ymin=0.5,
ymax=256,
xmin=1,
xmax=256,
legend style={at={(1,0.5)},
      anchor=west,legend columns=1},
xtick={1,2, 4,8, 16, 32, 64, 128, 256},
xticklabels={1,2, 4,8, 16, 32, 64, 128, 256},
ytick={1,2, 4,8, 16, 32, 64, 128, 256},
yticklabels={1,2, 4,8, 16, 32, 64, 128, 256},
xlabel={Number of Threads},
ylabel={Speedup},
title={Scalability of the Burgers Equation on KNL},
title style={font=\bfseries},
xmode=log,
ymode=log,
xmajorgrids,
ymajorgrids,
width=15em,
height=11.5em,
cycle list name=cbw,
scale only axis
]
\addplot table [y expr=\thisrow{priburgerS} / \thisrow{priburger}] {\MICTable}; \addlegendentry{Primal}
\addplot table [y expr=\thisrow{adjburgerS} / \thisrow{adjburger}] {\MICTable}; \addlegendentry{Adjoint}
\addplot table [y expr=\thisrow{atomicburgerS} / \thisrow{atomicburger}] {\MICTable}; \addlegendentry{Atomics}
\addplot table [y expr=\thisrow{perfburgerS} / \thisrow{perfburger}] {\MICTable}; \addlegendentry{PerforAD}
\addplot table [y expr=\thisrow{threads}] {\MICTable}; \addlegendentry{Ideal}
\end{axis}
\end{tikzpicture}
\caption{Speedups for the Burgers equation solver, showing near-perfect scalability up to 64 threads for the primal and adjoint stencil solver on a KNL processor. The scatter adjoints with atomics do not scale at all.}
\label{fig:scaleburger}
\end{figure}
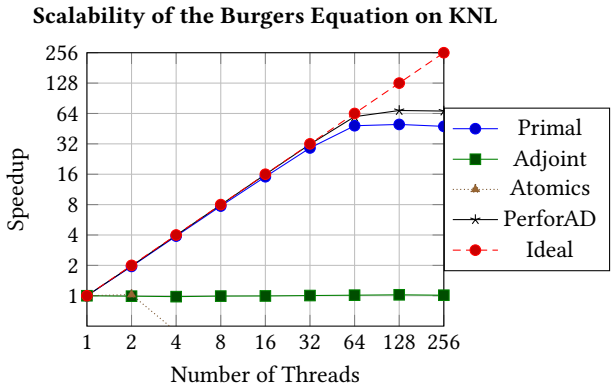

\begin{figure}
\begin{tikzpicture}
  \begin{axis}[
    ybar,
    ylabel={Runtime (s)},
    legend style={at={(0,1)},
      anchor=north west,legend columns=1},
    xtick={0,1,2,3,4}, xticklabels={Primal Serial, PerforAD Serial, Adjoint Serial, Primal Parallel, PerforAD Parallel },
    xtick=data,
    title={Runtimes of the Wave Equation on KNL},
title style={font=\bfseries},
ymajorgrids,
    nodes near coords, 
	nodes near coords align={above},
point meta=rawy,
    x tick label style={rotate=20,anchor=east},
width=15em,
height=11.5em,
	scale only axis,
	ymin=0.0,ymax=50,
	xmin=-0.5,xmax=4.5,
	clip=false
    ]
    \addplot[fill=black!70,draw=none] coordinates {
		(0, 12.817169)
		(1, 41.271231)
		(2, 25.450112) 
                (3, 0.843320)
		(4, 1.286821)
		};

\end{axis}
\end{tikzpicture}
\caption{Absolute runtimes for wave equation solvers. The fastest primal stencil used 128 threads, while the fastest adjoint stencil used 256 threads.Since the conventional adjoint code does not scale well, adjoint stencils lead to a much-reduced runtime in parallel, over $19\times$ faster than the best runtime of the conventional adjoint code.}
\label{fig:timewave}
\end{figure}
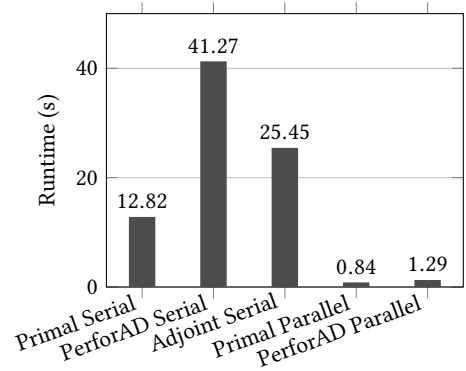

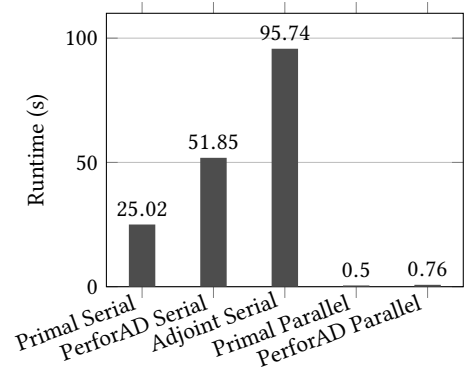
\begin{figure}
\begin{tikzpicture}
  \begin{axis}[
    ybar,
    ylabel={Runtime (s)},
    legend style={at={(0,1)},
      anchor=north west,legend columns=1},
    xtick={0,1,2,3,4}, xticklabels={Primal Serial, PerforAD Serial, Adjoint Serial, Primal Parallel, PerforAD Parallel },
    xtick=data,
    title={Runtimes Burgers Equation on KNL},
title style={font=\bfseries},
ymajorgrids,
    nodes near coords, 
	nodes near coords align={above},
point meta=rawy,
    x tick label style={rotate=20,anchor=east},
width=15em,
height=11.5em,
	scale only axis,
	ymin=0.0,ymax=110,
	xmin=-0.5,xmax=4.5,
	clip=false
    ]
    \addplot[fill=black!70,draw=none] coordinates {
		(0, 25.022697)
		(1, 51.853798)
		(2, 95.739452) 
                (3, 0.502064)
		(4, 0.756979)
		};

\end{axis}
\end{tikzpicture}
\caption{Absolute runtimes for the Burgers equation solvers. The adjoint stencil loop is faster than the conventional adjoint code generated by Tapenade. The latter is slowed by the access to the intermediate value stack that is used to store the results of the $\min$ and $\max$ functions. This, combined with the differences in scalability, results in a $125\times$ run time difference between Tapenade and \texttt{PerforAD}.}
\label{fig:timeburger}
\end{figure}

On the KNL system, the primal wave equation solver scales up to 16 threads, then plateaus. The solver cannot efficiently scale to the full number of threads, probably because the memory bandwidth is saturated. Adjoint stencils scale better than the primal stencils do (up to 32 threads), probably because the loop body contains more operations. The better scalability means that the performance gap between the primal and adjoint stencil is slightly smaller for parallel than for the serial execution.

The Tapenade wave equation adjoint is not parallelised and hence does not scale. The manually parallelised scatter adjoint with atomics scales up to 2 threads, then becomes slower with any additional thread. Its absolute runtime is always worse than that of the serial adjoint. The best-observed runtime for stencil adjoints is more than $19\times$ better than that of the conventional adjoint code.

For the Burgers equation, adjoint and primal stencils scale well up to 64 threads. Parallelised conventional adjoints with atomics always perform worse than serial conventional adjoints in our experiment. On KNL, we used the original Tapenade output that precomputes $\min$ and $\max$ functions on a stack for our serial run times. Hence, the \texttt{PerforAD}-generated adjoint stencil is faster even in serial. The manually parallelised Tapenade adjoint is identical to that used on Broadwell, with no stack access. Adjoint stencils outperform conventional adjoints by a factor of $125\times$.

\section{Conclusion and Future Work}
\label{sec:conclusion}

We have presented adjoint stencils, a method for automatic differentiation or back-propagation of gather stencil loops in a way that uses only gather stencil operations. This method is implemented in the open source tool \texttt{PerforAD}, which we release together with this paper. The test cases in our work demonstrate that adjoint stencils are as scalable on CPU and XeonPhi systems as are the original computations, and can outperform programs generated by conventional automatic differentiation by orders of magnitude.

We plan to test our method also on GPU systems, and to explore similar transformations not only for stencil computations but more generally for programs that have been performance-optimised. We aim to extend \texttt{PerforAD} by combining it with conventional automatic differentiation tools, as well as polyhedral compilers~\cite{uday08pldi,baghdadi2019tiramisu}, to target more applications. It would also be useful to integrate our tool with domain-specific language compilers~\cite{kamil2010auto,devito-compiler,Kronawitter:2018:PSS:3284745.3274653} to make these transformation available to a larger audience.

\noindent{\bf \\Acknowledgements}
This work was funded in part by 
support from the U.S. Department of Energy, 
Office of Science,  under contract DE-AC02-06CH11357. 
We gratefully acknowledge the computing resources provided on Blues, a high-performance computing cluster operated by the Laboratory Computing Resource Center at Argonne National Laboratory.

\bibliographystyle{ACM-Reference-Format}
\bibliography{stencilAD}


\begin{thebibliography}{27}


\ifx \showCODEN    \undefined \def \showCODEN     #1{\unskip}     \fi
\ifx \showDOI      \undefined \def \showDOI       #1{#1}\fi
\ifx \showISBNx    \undefined \def \showISBNx     #1{\unskip}     \fi
\ifx \showISBNxiii \undefined \def \showISBNxiii  #1{\unskip}     \fi
\ifx \showISSN     \undefined \def \showISSN      #1{\unskip}     \fi
\ifx \showLCCN     \undefined \def \showLCCN      #1{\unskip}     \fi
\ifx \shownote     \undefined \def \shownote      #1{#1}          \fi
\ifx \showarticletitle \undefined \def \showarticletitle #1{#1}   \fi
\ifx \showURL      \undefined \def \showURL       {\relax}        \fi
\providecommand\bibfield[2]{#2}
\providecommand\bibinfo[2]{#2}
\providecommand\natexlab[1]{#1}
\providecommand\showeprint[2][]{arXiv:#2}

\bibitem[\protect\citeauthoryear{{Araya-Polo}, {Cabezas}, {Hanzich}, {Pericas},
  {Rubio}, {Gelado}, {Shafiq}, {Morancho}, {Navarro}, {Ayguade}, {Cela}, and
  {Valero}}{{Araya-Polo} et~al\mbox{.}}{2011}]%
        {5628156}
\bibfield{author}{\bibinfo{person}{M. {Araya-Polo}}, \bibinfo{person}{J.
  {Cabezas}}, \bibinfo{person}{M. {Hanzich}}, \bibinfo{person}{M. {Pericas}},
  \bibinfo{person}{F. {Rubio}}, \bibinfo{person}{I. {Gelado}},
  \bibinfo{person}{M. {Shafiq}}, \bibinfo{person}{E. {Morancho}},
  \bibinfo{person}{N. {Navarro}}, \bibinfo{person}{E. {Ayguade}},
  \bibinfo{person}{J.~M. {Cela}}, {and} \bibinfo{person}{M. {Valero}}.}
  \bibinfo{year}{2011}\natexlab{}.
\newblock \showarticletitle{Assessing Accelerator-Based HPC Reverse Time
  Migration}.
\newblock \bibinfo{journal}{\emph{IEEE Transactions on Parallel and Distributed
  Systems}} \bibinfo{volume}{22}, \bibinfo{number}{1} (\bibinfo{date}{Jan}
  \bibinfo{year}{2011}), \bibinfo{pages}{147--162}.
\newblock
\showISSN{1045-9219}
\urldef\tempurl%
\url{https://doi.org/10.1109/TPDS.2010.144}
\showDOI{\tempurl}


\bibitem[\protect\citeauthoryear{Baghdadi, Ray, Romdhane, Del~Sozzo, Akkas,
  Zhang, Suriana, Kamil, and Amarasinghe}{Baghdadi et~al\mbox{.}}{2019}]%
        {baghdadi2019tiramisu}
\bibfield{author}{\bibinfo{person}{Riyadh Baghdadi}, \bibinfo{person}{Jessica
  Ray}, \bibinfo{person}{Malek~Ben Romdhane}, \bibinfo{person}{Emanuele
  Del~Sozzo}, \bibinfo{person}{Abdurrahman Akkas}, \bibinfo{person}{Yunming
  Zhang}, \bibinfo{person}{Patricia Suriana}, \bibinfo{person}{Shoaib Kamil},
  {and} \bibinfo{person}{Saman Amarasinghe}.} \bibinfo{year}{2019}\natexlab{}.
\newblock \showarticletitle{Tiramisu: A polyhedral compiler for expressing fast
  and portable code}. In \bibinfo{booktitle}{\emph{Proceedings of the 2019
  IEEE/ACM International Symposium on Code Generation and Optimization}}. IEEE
  Press, \bibinfo{pages}{193--205}.
\newblock


\bibitem[\protect\citeauthoryear{Baydin, Pearlmutter, Radul, and
  Siskind}{Baydin et~al\mbox{.}}{2018}]%
        {baydin2018automatic}
\bibfield{author}{\bibinfo{person}{Atilim~Gunes Baydin},
  \bibinfo{person}{Barak~A Pearlmutter}, \bibinfo{person}{Alexey~Andreyevich
  Radul}, {and} \bibinfo{person}{Jeffrey~Mark Siskind}.}
  \bibinfo{year}{2018}\natexlab{}.
\newblock \showarticletitle{Automatic differentiation in machine learning: a
  survey}.
\newblock \bibinfo{journal}{\emph{Journal of Marchine Learning Research}}
  \bibinfo{volume}{18} (\bibinfo{year}{2018}), \bibinfo{pages}{1--43}.
\newblock


\bibitem[\protect\citeauthoryear{Bondhugula, Hartono, Ramanujam, and
  Sadayappan}{Bondhugula et~al\mbox{.}}{2008}]%
        {uday08pldi}
\bibfield{author}{\bibinfo{person}{Uday Bondhugula}, \bibinfo{person}{Albert
  Hartono}, \bibinfo{person}{J. Ramanujam}, {and} \bibinfo{person}{P.
  Sadayappan}.} \bibinfo{year}{2008}\natexlab{}.
\newblock \showarticletitle{A practical automatic polyhedral program
  optimization system}. In \bibinfo{booktitle}{\emph{ACM SIGPLAN Conference on
  Programming Language Design and Implementation (PLDI)}}.
\newblock


\bibitem[\protect\citeauthoryear{Farrell, Ham, Funke, and Rognes}{Farrell
  et~al\mbox{.}}{2013}]%
        {farrell2013automated}
\bibfield{author}{\bibinfo{person}{Patrick~E Farrell}, \bibinfo{person}{David~A
  Ham}, \bibinfo{person}{Simon~W Funke}, {and} \bibinfo{person}{Marie~E
  Rognes}.} \bibinfo{year}{2013}\natexlab{}.
\newblock \showarticletitle{Automated derivation of the adjoint of high-level
  transient finite element programs}.
\newblock \bibinfo{journal}{\emph{SIAM Journal on Scientific Computing}}
  \bibinfo{volume}{35}, \bibinfo{number}{4} (\bibinfo{year}{2013}),
  \bibinfo{pages}{C369--C393}.
\newblock


\bibitem[\protect\citeauthoryear{F{\"o}rster}{F{\"o}rster}{2014}]%
        {forster2014algorithmic}
\bibfield{author}{\bibinfo{person}{Michael F{\"o}rster}.}
  \bibinfo{year}{2014}\natexlab{}.
\newblock \emph{\bibinfo{title}{Algorithmic Differentiation of Pragma-Defined
  Parallel Regions: Differentiating Computer Programs Containing {OpenMP}}}.
\newblock \bibinfo{thesistype}{Ph.D. Dissertation}. \bibinfo{school}{RWTH
  Aachen}.
\newblock


\bibitem[\protect\citeauthoryear{Giles, Ghate, and Duta}{Giles
  et~al\mbox{.}}{2005}]%
        {giles2005using}
\bibfield{author}{\bibinfo{person}{MB Giles}, \bibinfo{person}{D Ghate}, {and}
  \bibinfo{person}{MC Duta}.} \bibinfo{year}{2005}\natexlab{}.
\newblock \showarticletitle{Using automatic differentiation for adjoint {CFD}
  code development}.
\newblock  (\bibinfo{year}{2005}).
\newblock


\bibitem[\protect\citeauthoryear{Griewank et~al\mbox{.}}{Griewank
  et~al\mbox{.}}{1989}]%
        {griewank1989automatic}
\bibfield{author}{\bibinfo{person}{Andreas Griewank} {et~al\mbox{.}}}
  \bibinfo{year}{1989}\natexlab{}.
\newblock \showarticletitle{On automatic differentiation}.
\newblock \bibinfo{journal}{\emph{Mathematical Programming: recent developments
  and applications}} \bibinfo{volume}{6}, \bibinfo{number}{6}
  (\bibinfo{year}{1989}), \bibinfo{pages}{83--107}.
\newblock


\bibitem[\protect\citeauthoryear{Griewank, Juedes, and Utke}{Griewank
  et~al\mbox{.}}{1996}]%
        {griewank1996algorithm}
\bibfield{author}{\bibinfo{person}{Andreas Griewank}, \bibinfo{person}{David
  Juedes}, {and} \bibinfo{person}{Jean Utke}.} \bibinfo{year}{1996}\natexlab{}.
\newblock \showarticletitle{Algorithm 755: ADOL-C: a package for the automatic
  differentiation of algorithms written in C/C++}.
\newblock \bibinfo{journal}{\emph{ACM Transactions on Mathematical Software
  (TOMS)}} \bibinfo{volume}{22}, \bibinfo{number}{2} (\bibinfo{year}{1996}),
  \bibinfo{pages}{131--167}.
\newblock


\bibitem[\protect\citeauthoryear{Hascoet and Pascual}{Hascoet and
  Pascual}{2013}]%
        {tapenade}
\bibfield{author}{\bibinfo{person}{Laurent Hascoet} {and}
  \bibinfo{person}{Val{\'e}rie Pascual}.} \bibinfo{year}{2013}\natexlab{}.
\newblock \showarticletitle{The {T}apenade automatic differentiation tool:
  Principles, model, and specification}.
\newblock \bibinfo{journal}{\emph{ACM Trans. Math. Softw.}}
  \bibinfo{volume}{39}, \bibinfo{number}{3}, Article \bibinfo{articleno}{20}
  (\bibinfo{date}{May} \bibinfo{year}{2013}), \bibinfo{numpages}{43}~pages.
\newblock
\showISSN{0098-3500}
\urldef\tempurl%
\url{https://doi.org/10.1145/2450153.2450158}
\showDOI{\tempurl}


\bibitem[\protect\citeauthoryear{Heimbach, Hill, and Giering}{Heimbach
  et~al\mbox{.}}{2005}]%
        {heimbach2005efficient}
\bibfield{author}{\bibinfo{person}{Patrick Heimbach}, \bibinfo{person}{Chris
  Hill}, {and} \bibinfo{person}{Ralf Giering}.}
  \bibinfo{year}{2005}\natexlab{}.
\newblock \showarticletitle{An efficient exact adjoint of the parallel MIT
  general circulation model, generated via automatic differentiation}.
\newblock \bibinfo{journal}{\emph{Future Generation Computer Systems}}
  \bibinfo{volume}{21}, \bibinfo{number}{8} (\bibinfo{year}{2005}),
  \bibinfo{pages}{1356--1371}.
\newblock


\bibitem[\protect\citeauthoryear{Hogan}{Hogan}{2014}]%
        {hogan2014fast}
\bibfield{author}{\bibinfo{person}{Robin~J Hogan}.}
  \bibinfo{year}{2014}\natexlab{}.
\newblock \showarticletitle{Fast reverse-mode automatic differentiation using
  expression templates in C++}.
\newblock \bibinfo{journal}{\emph{ACM Transactions on Mathematical Software
  (TOMS)}} \bibinfo{volume}{40}, \bibinfo{number}{4} (\bibinfo{year}{2014}),
  \bibinfo{pages}{26}.
\newblock


\bibitem[\protect\citeauthoryear{Hovland}{Hovland}{1997}]%
        {hovland1997automatic}
\bibfield{author}{\bibinfo{person}{Paul~Dennis Hovland}.}
  \bibinfo{year}{1997}\natexlab{}.
\newblock \emph{\bibinfo{title}{Automatic differentiation of parallel
  programs}}.
\newblock \bibinfo{thesistype}{Ph.D. Dissertation}. \bibinfo{school}{University
  of Illinois at Urbana-Champaign}.
\newblock


\bibitem[\protect\citeauthoryear{H{\"u}ckelheim, Hovland, Strout, and
  M{\"u}ller}{H{\"u}ckelheim et~al\mbox{.}}{2018}]%
        {tfmad}
\bibfield{author}{\bibinfo{person}{J.C. H{\"u}ckelheim}, \bibinfo{person}{P.D.
  Hovland}, \bibinfo{person}{M.M. Strout}, {and} \bibinfo{person}{J.-D.
  M{\"u}ller}.} \bibinfo{year}{2018}\natexlab{}.
\newblock \showarticletitle{Parallelizable adjoint stencil computations using
  transposed forward-mode algorithmic differentiation}.
\newblock \bibinfo{journal}{\emph{Optimization Methods and Software}}
  \bibinfo{volume}{33}, \bibinfo{number}{4-6} (\bibinfo{year}{2018}),
  \bibinfo{pages}{672--693}.
\newblock
\urldef\tempurl%
\url{https://doi.org/10.1080/10556788.2018.1435654}
\showDOI{\tempurl}


\bibitem[\protect\citeauthoryear{H{\"u}ckelheim, Hovland, Strout, and
  M{\"u}ller}{H{\"u}ckelheim et~al\mbox{.}}{2017}]%
        {ssmp}
\bibfield{author}{\bibinfo{person}{Jan H{\"u}ckelheim},
  \bibinfo{person}{Paul~D. Hovland}, \bibinfo{person}{Michelle~Mills Strout},
  {and} \bibinfo{person}{Jens-Dominik M{\"u}ller}.}
  \bibinfo{year}{2017}\natexlab{}.
\newblock \showarticletitle{Reverse-mode algorithmic differentiation of an
  {OpenMP}-parallel compressible flow solver}.
\newblock \bibinfo{journal}{\emph{International Journal for High Performance
  Computing Applications}} (\bibinfo{year}{2017}).
\newblock
\urldef\tempurl%
\url{https://doi.org/10.1177/1094342017712060}
\showDOI{\tempurl}


\bibitem[\protect\citeauthoryear{Innes}{Innes}{2018}]%
        {innes2018don}
\bibfield{author}{\bibinfo{person}{Michael Innes}.}
  \bibinfo{year}{2018}\natexlab{}.
\newblock \showarticletitle{Don't unroll adjoint: differentiating SSA-Form
  programs}.
\newblock \bibinfo{journal}{\emph{arXiv preprint arXiv:1810.07951}}
  (\bibinfo{year}{2018}).
\newblock


\bibitem[\protect\citeauthoryear{Kamil, Chan, Oliker, Shalf, and
  Williams}{Kamil et~al\mbox{.}}{2010}]%
        {kamil2010auto}
\bibfield{author}{\bibinfo{person}{Shoaib Kamil}, \bibinfo{person}{Cy Chan},
  \bibinfo{person}{Leonid Oliker}, \bibinfo{person}{John Shalf}, {and}
  \bibinfo{person}{Samuel Williams}.} \bibinfo{year}{2010}\natexlab{}.
\newblock \showarticletitle{An auto-tuning framework for parallel multicore
  stencil computations}. In \bibinfo{booktitle}{\emph{2010 IEEE International
  Symposium on Parallel \& Distributed Processing (IPDPS)}}. IEEE,
  \bibinfo{pages}{1--12}.
\newblock


\bibitem[\protect\citeauthoryear{Kronawitter and Lengauer}{Kronawitter and
  Lengauer}{2018}]%
        {Kronawitter:2018:PSS:3284745.3274653}
\bibfield{author}{\bibinfo{person}{Stefan Kronawitter} {and}
  \bibinfo{person}{Christian Lengauer}.} \bibinfo{year}{2018}\natexlab{}.
\newblock \showarticletitle{Polyhedral search space exploration in the
  ExaStencils code generator}.
\newblock \bibinfo{journal}{\emph{ACM Trans. Archit. Code Optim.}}
  \bibinfo{volume}{15}, \bibinfo{number}{4}, Article \bibinfo{articleno}{40}
  (\bibinfo{date}{Oct.} \bibinfo{year}{2018}), \bibinfo{numpages}{25}~pages.
\newblock
\showISSN{1544-3566}
\urldef\tempurl%
\url{https://doi.org/10.1145/3274653}
\showDOI{\tempurl}


\bibitem[\protect\citeauthoryear{Li, Gharbi, Adams, Durand, and
  Ragan-Kelley}{Li et~al\mbox{.}}{2018}]%
        {li2018differentiable}
\bibfield{author}{\bibinfo{person}{Tzu-Mao Li}, \bibinfo{person}{Micha{\"e}l
  Gharbi}, \bibinfo{person}{Andrew Adams}, \bibinfo{person}{Fr{\'e}do Durand},
  {and} \bibinfo{person}{Jonathan Ragan-Kelley}.}
  \bibinfo{year}{2018}\natexlab{}.
\newblock \showarticletitle{Differentiable programming for image processing and
  deep learning in Halide}.
\newblock \bibinfo{journal}{\emph{ACM Transactions on Graphics (TOG)}}
  \bibinfo{volume}{37}, \bibinfo{number}{4} (\bibinfo{year}{2018}),
  \bibinfo{pages}{139}.
\newblock


\bibitem[\protect\citeauthoryear{{Luporini}, {Lange}, {Louboutin}, {Kukreja},
  {H{\"u}ckelheim}, {Yount}, {Witte}, {Kelly}, {Gorman}, and
  {Herrmann}}{{Luporini} et~al\mbox{.}}{2018}]%
        {devito-compiler}
\bibfield{author}{\bibinfo{person}{F. {Luporini}}, \bibinfo{person}{M.
  {Lange}}, \bibinfo{person}{M. {Louboutin}}, \bibinfo{person}{N. {Kukreja}},
  \bibinfo{person}{J. {H{\"u}ckelheim}}, \bibinfo{person}{C. {Yount}},
  \bibinfo{person}{P. {Witte}}, \bibinfo{person}{P.~H.~J. {Kelly}},
  \bibinfo{person}{G.~J. {Gorman}}, {and} \bibinfo{person}{F.~J. {Herrmann}}.}
  \bibinfo{year}{2018}\natexlab{}.
\newblock \showarticletitle{Architecture and performance of Devito, a system
  for automated stencil computation}.
\newblock \bibinfo{journal}{\emph{CoRR}}  \bibinfo{volume}{abs/1807.03032}
  (\bibinfo{date}{jul} \bibinfo{year}{2018}).
\newblock
\showeprint[arxiv]{1807.03032}
\urldef\tempurl%
\url{http://arxiv.org/abs/1807.03032}
\showURL{%
\tempurl}


\bibitem[\protect\citeauthoryear{Meurer, Smith, Paprocki, \v{C}ert\'{i}k,
  Kirpichev, Rocklin, Kumar, Ivanov, Moore, Singh, Rathnayake, Vig, Granger,
  Muller, Bonazzi, Gupta, Vats, Johansson, Pedregosa, Curry, Terrel,
  Rou\v{c}ka, Saboo, Fernando, Kulal, Cimrman, and Scopatz}{Meurer
  et~al\mbox{.}}{2017}]%
        {10.7717/peerj-cs.103}
\bibfield{author}{\bibinfo{person}{Aaron Meurer},
  \bibinfo{person}{Christopher~P. Smith}, \bibinfo{person}{Mateusz Paprocki},
  \bibinfo{person}{Ond\v{r}ej \v{C}ert\'{i}k}, \bibinfo{person}{Sergey~B.
  Kirpichev}, \bibinfo{person}{Matthew Rocklin}, \bibinfo{person}{AMiT Kumar},
  \bibinfo{person}{Sergiu Ivanov}, \bibinfo{person}{Jason~K. Moore},
  \bibinfo{person}{Sartaj Singh}, \bibinfo{person}{Thilina Rathnayake},
  \bibinfo{person}{Sean Vig}, \bibinfo{person}{Brian~E. Granger},
  \bibinfo{person}{Richard~P. Muller}, \bibinfo{person}{Francesco Bonazzi},
  \bibinfo{person}{Harsh Gupta}, \bibinfo{person}{Shivam Vats},
  \bibinfo{person}{Fredrik Johansson}, \bibinfo{person}{Fabian Pedregosa},
  \bibinfo{person}{Matthew~J. Curry}, \bibinfo{person}{Andy~R. Terrel},
  \bibinfo{person}{\v{S}t\v{e}p\'{a}n Rou\v{c}ka}, \bibinfo{person}{Ashutosh
  Saboo}, \bibinfo{person}{Isuru Fernando}, \bibinfo{person}{Sumith Kulal},
  \bibinfo{person}{Robert Cimrman}, {and} \bibinfo{person}{Anthony Scopatz}.}
  \bibinfo{year}{2017}\natexlab{}.
\newblock \showarticletitle{SymPy: symbolic computing in Python}.
\newblock \bibinfo{journal}{\emph{PeerJ Computer Science}}  \bibinfo{volume}{3}
  (\bibinfo{date}{Jan.} \bibinfo{year}{2017}), \bibinfo{pages}{e103}.
\newblock
\showISSN{2376-5992}
\urldef\tempurl%
\url{https://doi.org/10.7717/peerj-cs.103}
\showDOI{\tempurl}


\bibitem[\protect\citeauthoryear{Narayanan, Norris, and Winnicka}{Narayanan
  et~al\mbox{.}}{2010}]%
        {narayanan2010adic2}
\bibfield{author}{\bibinfo{person}{Sri Hari~Krishna Narayanan},
  \bibinfo{person}{Boyana Norris}, {and} \bibinfo{person}{Beata Winnicka}.}
  \bibinfo{year}{2010}\natexlab{}.
\newblock \showarticletitle{ADIC2: Development of a component source
  transformation system for differentiating C and C++}.
\newblock \bibinfo{journal}{\emph{Procedia Computer Science}}
  \bibinfo{volume}{1}, \bibinfo{number}{1} (\bibinfo{year}{2010}),
  \bibinfo{pages}{1845--1853}.
\newblock


\bibitem[\protect\citeauthoryear{Paszke, Gross, Chintala, Chanan, Yang, DeVito,
  Lin, Desmaison, Antiga, and Lerer}{Paszke et~al\mbox{.}}{2017}]%
        {paszke2017automatic}
\bibfield{author}{\bibinfo{person}{Adam Paszke}, \bibinfo{person}{Sam Gross},
  \bibinfo{person}{Soumith Chintala}, \bibinfo{person}{Gregory Chanan},
  \bibinfo{person}{Edward Yang}, \bibinfo{person}{Zachary DeVito},
  \bibinfo{person}{Zeming Lin}, \bibinfo{person}{Alban Desmaison},
  \bibinfo{person}{Luca Antiga}, {and} \bibinfo{person}{Adam Lerer}.}
  \bibinfo{year}{2017}\natexlab{}.
\newblock \showarticletitle{Automatic differentiation in PyTorch}.
\newblock  (\bibinfo{year}{2017}).
\newblock


\bibitem[\protect\citeauthoryear{Ragan-Kelley, Barnes, Adams, Paris, Durand,
  and Amarasinghe}{Ragan-Kelley et~al\mbox{.}}{2013}]%
        {ragan2013halide}
\bibfield{author}{\bibinfo{person}{Jonathan Ragan-Kelley},
  \bibinfo{person}{Connelly Barnes}, \bibinfo{person}{Andrew Adams},
  \bibinfo{person}{Sylvain Paris}, \bibinfo{person}{Fr{\'e}do Durand}, {and}
  \bibinfo{person}{Saman Amarasinghe}.} \bibinfo{year}{2013}\natexlab{}.
\newblock \showarticletitle{Halide: A language and compiler for optimizing
  parallelism, locality, and recomputation in image processing pipelines}.
\newblock \bibinfo{journal}{\emph{ACM SIGPLAN Notices}} \bibinfo{volume}{48},
  \bibinfo{number}{6} (\bibinfo{year}{2013}), \bibinfo{pages}{519--530}.
\newblock


\bibitem[\protect\citeauthoryear{Revels, Lubin, and Papamarkou}{Revels
  et~al\mbox{.}}{2016}]%
        {revels2016forward}
\bibfield{author}{\bibinfo{person}{Jarrett Revels}, \bibinfo{person}{Miles
  Lubin}, {and} \bibinfo{person}{Theodore Papamarkou}.}
  \bibinfo{year}{2016}\natexlab{}.
\newblock \showarticletitle{Forward-mode automatic differentiation in Julia}.
\newblock \bibinfo{journal}{\emph{arXiv preprint arXiv:1607.07892}}
  (\bibinfo{year}{2016}).
\newblock


\bibitem[\protect\citeauthoryear{Stock, Kong, Grosser, Pouchet, Rastello,
  Ramanujam, and Sadayappan}{Stock et~al\mbox{.}}{2014}]%
        {stock2014framework}
\bibfield{author}{\bibinfo{person}{Kevin Stock}, \bibinfo{person}{Martin Kong},
  \bibinfo{person}{Tobias Grosser}, \bibinfo{person}{Louis-No{\"e}l Pouchet},
  \bibinfo{person}{Fabrice Rastello}, \bibinfo{person}{Jagannathan Ramanujam},
  {and} \bibinfo{person}{Ponnuswamy Sadayappan}.}
  \bibinfo{year}{2014}\natexlab{}.
\newblock \showarticletitle{A framework for enhancing data reuse via
  associative reordering}. In \bibinfo{booktitle}{\emph{ACM SIGPLAN Notices}},
  Vol.~\bibinfo{volume}{49}. ACM, \bibinfo{pages}{65--76}.
\newblock


\bibitem[\protect\citeauthoryear{Utke, Naumann, Fagan, Tallent, Strout,
  Heimbach, Hill, and Wunsch}{Utke et~al\mbox{.}}{2008}]%
        {utke2008openad}
\bibfield{author}{\bibinfo{person}{Jean Utke}, \bibinfo{person}{Uwe Naumann},
  \bibinfo{person}{Mike Fagan}, \bibinfo{person}{Nathan Tallent},
  \bibinfo{person}{Michelle Strout}, \bibinfo{person}{Patrick Heimbach},
  \bibinfo{person}{Chris Hill}, {and} \bibinfo{person}{Carl Wunsch}.}
  \bibinfo{year}{2008}\natexlab{}.
\newblock \showarticletitle{OpenAD/F: A modular open-source tool for automatic
  differentiation of Fortran codes}.
\newblock \bibinfo{journal}{\emph{ACM Transactions on Mathematical Software
  (TOMS)}} \bibinfo{volume}{34}, \bibinfo{number}{4} (\bibinfo{year}{2008}),
  \bibinfo{pages}{18}.
\newblock


\end{thebibliography}

\vfill
\begin{flushright}
\tiny
\framebox{\parbox{3.0in}{The submitted manuscript has been created by UChicago Argonne, LLC, Operator of Argonne National Laboratory (`Argonne'). Argonne, a U.S. Department of Energy Office of Science laboratory, is operated under Contract No. DE-AC02-06CH11357. The U.S. Government retains for itself, and others acting on its behalf, a paid-up nonexclusive, irrevocable worldwide license in said article to reproduce, prepare derivative works, distribute copies to the public, and perform publicly and display publicly, by or on behalf of the Government.  The Department of Energy will provide public access to these results of federally sponsored research in accordance with the DOE Public Access Plan. \url{http://energy.gov/downloads/doe-public-access-plan}.}}
\normalsize
\end{flushright}
\end{document}